\documentclass[preprint]{aastex}
\usepackage{psfig}

\slugcomment{KSUPT-01/4 \hspace{0.5truecm} June 2001}

\newcommand{\omegal}{\Omega_{\Lambda}}

\begin{document}

\title{Cosmological Parameter Determination from Counts of Galaxies}

\author{Silviu Podariu and Bharat Ratra}

\affil{Department of Physics, Kansas State University, Manhattan, KS 66506.}

\begin{abstract}
We study constraints that anticipated DEEP survey galaxy counts versus 
redshift data will place on cosmological model parameters in models with 
and without a constant or time-variable cosmological constant $\Lambda$. 
This data will result in fairly tight constraints on these parameters. 
For example, if all other parameters of a spatially-flat model with a 
constant $\Lambda$ are known, the galaxy counts data should constrain the 
nonrelativistic matter density parameter $\Omega_0$ to about 5\% 
(10\%, 1.5\%) at 1 $\sigma$ with neutral (worst case, best case) assumptions 
about data quality.
\end{abstract}

\keywords{cosmological parameters---cosmology: observation---large-scale 
structure of the universe---galaxies: general}

\section{Introduction} 

Current observational data favors cosmogonies with a low $\Omega_0$. The 
simplest such cold dark matter models have either flat spatial hypersurfaces
and a constant or time-variable cosmological ``constant" $\Lambda$ (see, e.g., 
Peebles 1984; Peebles \& Ratra 1988, hereafter PR; Ratra et al. 1997; Sahni 
\& Starobinsky 2000; Steinhardt 1999; Carroll 2001), or open spatial 
hypersurfaces and no $\Lambda$ (see, e.g., Gott 1982, 1997; Ratra \& Peebles 
1994, 1995; Cole et al. 1997; G\'orski et al.  1998). For a constant $\Lambda$ 
with density parameter $\omegal$, these models lie along the lines $\Omega_0 + 
\omegal = 1$ and $\omegal = 0$, respectively, in the two-dimensional 
($\Omega_0$, $\omegal$) parameter space. Models in this two-dimensional parameter space have either closed, flat, or open spatial hypersurfaces, 
depending on the values of $\Omega_0$ and $\omegal$. In this paper we study the 
general two-dimensional model as well as the special one-dimensional cases.

We also study a spatially-flat model with a time-variable $\Lambda$. The only 
known consistent realization of this quintessence scenario is that based on a 
scalar field ($\phi$) with a scalar field potential $V(\phi)$ (Ratra \& 
Peebles 1988). In this paper we focus on the favored model which at low 
redshift $z$ has $V(\phi) \propto \phi^{-\alpha}$, $\alpha > 0$ (PR; Ratra \& 
Peebles 1988).\footnote{For recent discussions of quintessence see, e.g., Weinberg (2000), Matos \& Ure\~na-L\'opez (2001), Armendariz-Picon, Mukhanov, 
\& Steinhardt (2001), McDonald (2001), de Ritis \& Marino (2000), Kaganovich 
(2001), Bean \& Magueijo (2000), Sen \& Seshadri (2000), Nunes \& Mimoso (2000),
and Hebecker \& Wetterich (2001). Inverse power law scalar field potentials 
appear in some high energy particle physics models (see, e.g., Rosati 2001; 
Brax, Martin, \& Riazuelo 2001). Brane quintessence models have also been 
discussed by, e.g., Maeda (2000), Huey \& Lidsey (2001), Albrecht et al. (2001),
Brax \& Davis (2001), and Majumdar (2001). However, it appears non-trivial for 
string/M theory to accommodate quintessence (see, e.g., Hellerman, Kaloper, \& 
Susskind 2001; Fischler et al. 2001; Moffat 2001; Halyo 2001; Cline 2001).}
A scalar field is mathematically equivalent to a fluid with a
time-dependent speed of sound (Ratra 1991), and with $V(\phi) \propto 
\phi^{-\alpha}$, $\alpha > 0$, the $\phi$ energy density behaves like a 
cosmological constant that decreases with time. In our analysis of this model 
here we do not make use of the time-independent equation of state fluid 
approximation to the model, since this leads to incorrect results (e.g., 
Podariu \& Ratra 2000). Effective fluid realizations of quintessence with a 
time-dependent equation of state have also been studied. However, since 
such an equation of state is arbitrary, observational constraints on it 
are largely determined by how it is modelled. Scalar field realizations
of quintessence are much more compelling, and only a handful of simple 
$V(\phi)$'s exhibit quintessence.

Cosmic microwave background anisotropy measurements have been used to
discriminate between the flat and open models (see, e.g., Ratra et al. 1999;
Rocha et al. 1999; Knox \& Page 2000; Douspis et al. 2001; Podariu et al. 
2001b; Netterfield et al. 2001; Pryke et al. 2001; Stompor et al. 2001),
and favor the flat case. These observations have also been used to constrain 
quintessence models, see, e.g., Brax, Martin, \& Riazuelo (2000), Amendola 
(2001), Balbi et al. (2001), Doran et al. (2001), Gonz\'alez-D\'{\i}az (2001), 
and Schulz \& White (2001). 

The flat-constant-$\Lambda$ model seems to be in conflict with a number of 
observations, including: (1) analyses of the rate of gravitational lensing of 
quasars and radio sources by foreground galaxies (see, e.g., Falco, Kochanek, 
\& Mu\~noz 1998); and (2) analyses of the number of large arcs formed by 
strong gravitational lensing by clusters (Bartelmann et al. 1998).
A spatially-flat quintessence model can accommodate the first constraint.
See Ratra \& Quillen (1992), Frieman \& Waga (1998), and Waga \& Frieman (2000)
for discussions of the scalar field quintessence case, and Zhu (2000), Cappi 
(2001), and Dev et al. (2001) for the fluid quintessence case.

Type Ia supernova (SN~Ia) apparent magnitude versus redshift data favor the 
flat model (see, e.g., Riess et al. 1998, 2001; Perlmutter et al. 1999; Podariu 
\& Ratra 2000; Gott et al. 2001). For SN Ia constraints on scalar field 
quintessence models see, e.g., Podariu \& Ratra (2000), Waga \& Frieman (2000), 
Gott et al. (2001), Wiltshire (2001), and Pavlov et al. (2001). Turner \& Riess 
(2001) discuss the effective fluid quintessence case. Higher quality SN Ia data,
such as that anticipated from the proposed SNAP space 
telescope\footnote{http://snap.lbl.gov/}, will result in tighter constraints on 
cosmological parameters. See Podariu, Nugent, \& Ratra (2001b, hereafter PNR) 
for a discussion of the scalar field quintessence case, and e.g., Maor, 
Brustein, \& Steinhardt (2001), Chevallier \& Polarski (2001), Barger \& 
Marfatia (2001), Huterer \& Turner (2000), Wang \& Garnavich (2001), Goliath 
et al. (2001), and Weller \& Albrecht (2001) for discussions of the effective 
fluid quintessence case. 

Loh \& Spillar (1986) applied the number counts versus redshift test (Peebles 
1993, $\S$13) to a set of galaxies with photometric redshifts. For the redshift 
range of this test, galaxies are assumed to be conserved and the shape of the
luminosity function of these galaxies is assumed to not change. Newman \&
Davis (2000, hereafter ND) suggest that anticipated DEEP (Deep Extragalactic
Evolutionary Probe)\footnote{http://deep.ucolick.org/} survey data on the number 
of galaxies (halos), at fixed rotation speed\footnote{ND argue that at fixed 
rotation speed the abundance of halos is almost independent of cosmological 
model and may be calibrated using semianalytical or numerical models.}, 
as a function of redshift will be an ideal candidate for the number counts
test.\footnote{Haiman, Mohr, \& Holder (2001) discuss the prospects of using 
X-ray and Sunyaev-Zeldovich effect selected clusters for this test.} 
ND examine constraints from anticipated DEEP data on the parameters of the 
general constant $\Lambda$ two-dimensional ($\Omega_0$, $\omegal$) model. They 
as well as Maor et al. (2001) and Huterer \& Turner (2000) also study 
constraints on effective fluid quintessence models.

In this paper we focus on how well anticipated DEEP data will constrain 
parameters of various cosmological models. More specifically, for the 
quintessence case, for reasons discussed above, we focus on the favored scalar 
field model with $V(\phi) \propto \phi^{-\alpha}$, $\alpha > 0$ (PR; Ratra \& 
Peebles 1988) instead of the effective fluid models considered by ND, Maor 
et al. (2001), and Huterer \& Turner (2000).

We want to determine how well anticipated DEEP data discriminates between
different cosmological model parameter values. To do this we pick a model 
and a range of model-parameter values and compute the predicted count of
objects per steradian and per unit redshift increment, $dN/dz(z)$, for a grid 
of model parameter values that span this range. Figure 1 shows examples of 
$(H_0{}^3/n_0) dN/dz(z)$'s (here $H_0$ is the Hubble constant and $n_0$ is
the proper number density of objects at $z = 0$) computed in the time-variable 
$\Lambda$ model (PR).

We follow ND and assume that number counts data from DEEP will be combined to 
provide $dN/dz(z)$'s and errors on $dN/dz(z)$'s for 8 uniform bins in redshift 
between $z = 0.7$ and $z = 1.5$. In each redshift bin the statistical and 
systematic errors are combined to give a $dN/dz(z)$ error distribution with 
standard deviation $\sigma (z)$. ND consider 10,000 galaxies distributed in 
redshift as $(1 + z)^{-2}$. The ``best" case estimate of $\sigma (z)$ assumes 
Poisson errors only (ND) and results in $\sigma (z_i)$ = 2.40, 2.54, 2.67, 2.81,
2.95, 3.09, 3.22, 3.36 \% for bins centered at $z_i$ = 0.75, 0.85, \dots, 
1.45, respectively.\footnote{Since even the last bin contains a large number 
of galaxies, about 890, the Poisson distribution is close to Gaussian.}
To account for the uncertainty due to evolution, Huterer \& Turner (2000)
consider ``neutral" (``worst") case estimates of $\sigma (z)$ determined by
adding 10\% (20\%) in quadrature to the best case $\sigma (z_i)$ values.

To determine how well number counts data will discriminate between different 
sets of parameter values, we pick a fiducial set of parameter values which give 
$dN_{\rm F}/dz(z)$ and compute
\begin{equation}
  N_\sigma (P) = \sqrt{ \sum_{i = 1}^{8} \left({dN/dz(P, z_i) - dN_{\rm F}/dz
      (z_i) \over \sigma(z_i) dN_{\rm F}/dz (z_i) } \right)^2 } ,
\end{equation}
where the sum runs over the 8 redshift bins and $P$ represents the
model parameters, for instance $\Omega_0$ and $\omegal$ in the general
two-dimensional constant $\Lambda$ case. $N_\sigma (P)$ is the number
of standard deviations the parameter set $P$ lies away from that
of the fiducial model. 

Results are presented and discussed in $\S$2 and we conclude in $\S$3.

\section{Results and Discussion} 

As a test of our method, we compute constraints for two models ND study, the 
$\Omega_0 = 0.3$, $\omegal = 0.7$ constant $\Lambda$ spatially-flat model in 
the upper panel of their Fig. 3 and the $\Omega_0 = 0.3$, fluid equation of 
state parameter $w = -1$ spatially-flat effective fluid quintessence model 
in the lower panel of their Fig. 3 (and Fig. 14 of Huterer \& Turner 2000). 
In both cases our constraint contours are in very good agreement with those 
derived by ND (and by Huterer \& Turner 2000).

Figure 2 illustrates the ability of anticipated DEEP data to constrain
cosmological parameters ($\Omega_0$ and $\omegal$) for the general 
two-dimensional constant $\Lambda$ case. The chosen fiducial model is 
spatially-flat with $\Omega_0 = 0.28$ and $\omegal = 0.72$. As expected, 
the constraint contours are elliptical, indicating that one combination 
of the parameters is better constrained than the other orthogonal 
combination. DEEP data with neutral case errors will lead to interesting 
constraints on cosmological parameters. DEEP data with best case errors 
will result in tighter constraints than current SN Ia data (compare with 
Fig. 5 of Podariu \& Ratra 2000) but will be slightly less constraining 
than anticipated worst case SNAP data (see Fig. 3 of PNR). Note however that 
the worst case SNAP data errors of PNR are the baseline SNAP mission errors.

ND note that the major axis of the anticipated DEEP data constraint contours 
is rotated slightly relative to that of the anticipated SNAP data contours 
(compare Fig. 2 here with Fig. 3 of PNR). If DEEP data systematic errors 
(largely due to evolution) can be reduced well below the Huterer \& Turner 
(2000) estimates, then a combined analysis of anticipated DEEP and SNAP data 
will result in very tight constraints on cosmological parameters. If the DEEP 
data systematic errors remain as large as the Huterer \& Turner (2000) 
estimates, then this data will be less constraining than SNAP data and the 
slight relative rotation of the major axes of the contours will result in 
DEEP data only slightly tightening the SNAP contours. 

Figure 3 illustrates the ability of anticipated DEEP data to discriminate 
between a constant and a time-variable $\Lambda$ in a spatially-flat model. 
The chosen fiducial model has $\Omega_0 = 0.28$ and $\alpha = 0$, and is a 
constant $\Lambda$ model with $\omegal = 0.72$. Again, as expected, the 
contours are elliptical. DEEP data with best case errors will lead to good 
discrimination, roughly comparable to what should be achieved by anticipated 
worst case SNAP data (see Fig. 4 of PNR).

Figure 4 illustrates the ability of DEEP data to constrain $\Omega_0$ and 
$\alpha$ in the spatially-flat time-variable $\Lambda$ model (PR). The 
chosen fiducial model has $\Omega_0 = 0.2$ and $\alpha = 4$. Again, 
anticipated DEEP data with best case errors will result in tight constraints, 
roughly comparable to what should be achieved by anticipated worst case SNAP 
data (see Fig. 5 of PNR).

If other data (e.g., cosmic microwave background anisotropy, or
weak-lensing, or SN Ia apparent magnitude measurements) pinned down some of 
the cosmological parameters, DEEP data would then provide tighter
constraints on the remaining parameters. For example, Fig. 5 shows
constraints from anticipated DEEP data on $\Omega_0$ in a spatially-flat 
constant $\Lambda$ model and in an open $\Lambda = 0$ model. DEEP data will
provide fairly restrictive constraints on $\Omega_0$ in both cases. For 
instance, at 3 $\sigma$, in the spatially-flat model (with $\Omega_0 = 0.3$ 
and $\omegal = 0.7$) we find $\Omega_0 = 0.3 {+0.05 \atop -0.04}$, 
$= 0.3 {+0.11 \atop -0.07}$, and $= 0.3 {+0.01 \atop -0.01}$ for neutral, 
worst, and best case errors, while in the open model (with $\Omega_0 = 0.3$ 
and $\omegal = 0$) we have $\Omega_0 = 0.3 {+0.12 \atop -0.10}$, $= 0.3 
{+0.26 \atop -0.18}$, and $= 0.3 {+0.03 \atop - 0.03}$ for neutral, worst, 
and best case errors. As expected from the elliptical shape of the contours 
in Fig. 2, anticipated DEEP data will constrain $\Omega_0$ more tightly in 
the flat case than in the open case. 

Figure 6 shows DEEP data constraints on $\Omega_0$ and $\alpha$ in the 
spatially flat time-variable $\Lambda$ model, if other data were to require 
that either $\alpha = 4$ or $\Omega_0 = 0.2$ (in a fiducial model with 
$\Omega_0 = 0.2$ and $\alpha = 4$). Anticipated DEEP data will 
provide fairly tight constraints on these parameters. For example, if $\alpha 
= 4$ we find $\Omega_0 = 0.2 {+0.07 \atop -0.05}$, $= 0.2 {+0.15 \atop -0.08}$, 
and $= 0.2 {+0.02 \atop -0.01}$ for neutral, worst, and best case errors, while 
if $\Omega_0 = 0.2$ we have $\alpha = 4 {+2.2 \atop -1.3}$, $= 4 {+(>4) \atop
-2.1}$, and $= 4 {+0.5 \atop -0.4}$ for neutral, worst (here the upper limit 
lies outside the parameter range considered), and best case errors, all at 3 
$\sigma$.

\section{Conclusion}

Galaxy number counts versus redshift data of the quality assumed here will 
lead to fairly tight constraints on cosmo\-logical parameters. For example,
in a spatially-flat constant $\Lambda$ model where all other parameters are 
known, anticipated DEEP data will determine $\Omega_0$ to about $\pm 5.1\%$, 
$\pm 9.8\%$, and $\pm 1.4\%$ (for neutral, worst, and best case errors 
respectively) at 1 $\sigma$. The corresponding errors on $\Omega_0$ for the 
open case are about $\pm 12\%$, $\pm 24\%$, and $\pm 3.4\%$. For the 
time-variable $\Lambda$ model, when $\alpha$ is fixed, $\Omega_0$ will be 
known to about $\pm 9.2\%$, $\pm 18\%$, and $\pm 2.6\%$, respectively, while 
when $\Omega_0$ is fixed, $\alpha$ will be determined to about $\pm 14\%$,
$\pm 27\%$, and $\pm 3.8\%$. In agreement with ND, we find that anticipated 
DEEP data with best case errors will be roughly as constraining as anticipated 
SNAP data with worst case errors, i.e., SNAP baseline mission errors (see 
$\S$4 of PNR). 

\bigskip

We acknowledge useful discussions with M. Davis, D. Huterer, P. Mukherjee, and 
P. Nugent, and support from NSF CAREER grant AST-9875031. 

\bigskip

\begin{figure}[p]
\psfig{file=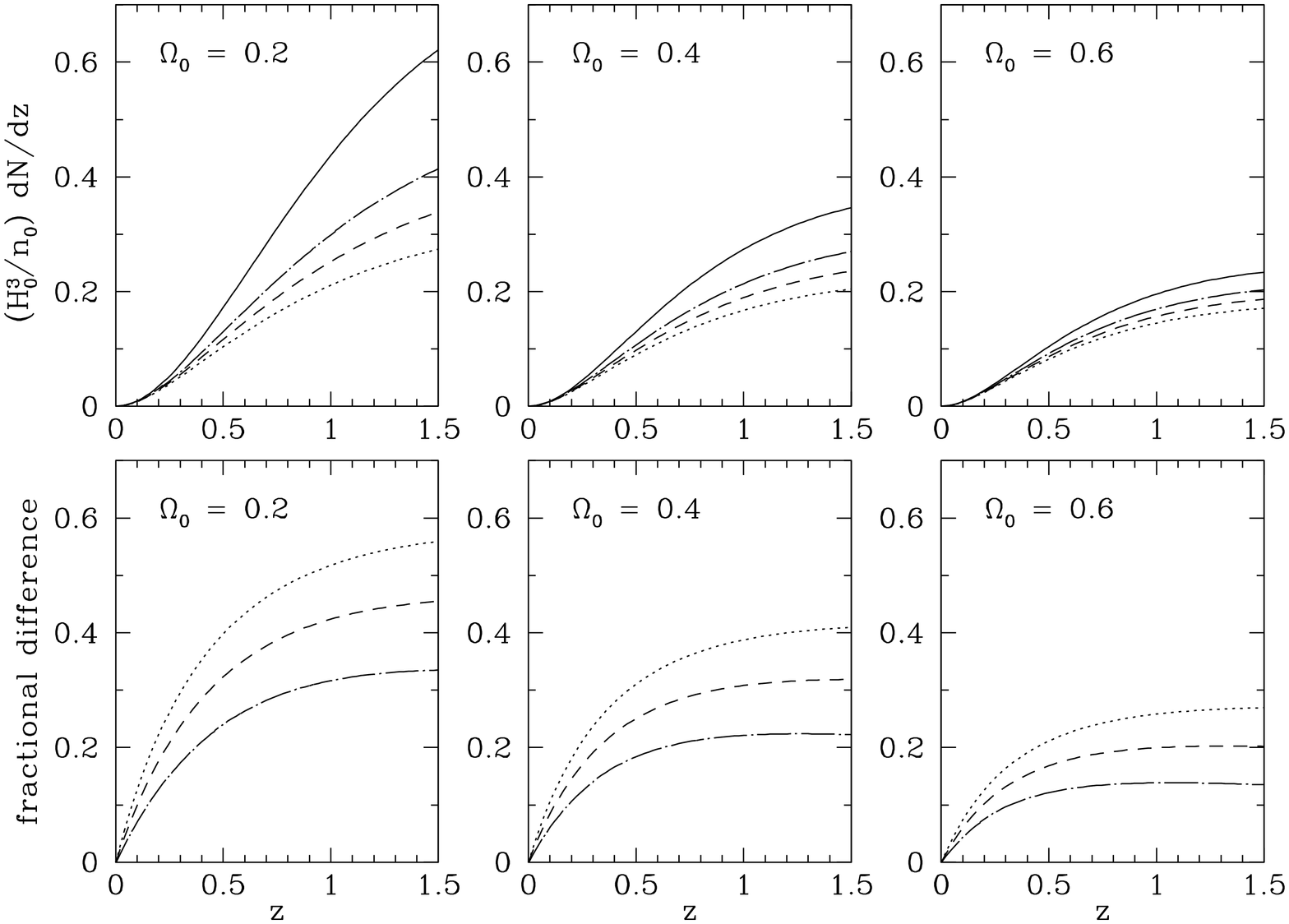,height=7.0in,width=7.0in,angle=270}
\caption{Lines in the panels in the upper row show normalized counts of 
objects per steradian and per unit redshift interval $(H_0{}^3/n_0) dN/dz (z, 
\alpha)$ as a function of redshift $z$ for various values of $\alpha$ in the 
spatially-flat time-variable $\Lambda$ model with scalar field potential 
$V(\phi) \propto \phi^{-\alpha}$. In descending order at $z = 1.5$ the lines 
correspond to $\alpha$ = 0, 2, 4, and 8 (solid, dot-dashed, dashed, and 
dotted curves respectively). $\alpha = 0$ is the constant $\Lambda$ model. 
From left to right the three panels correspond to $\Omega_0$ = 0.2, 0.4, and 
0.6.  The three lower panels show the fractional differences relative to the 
$\alpha = 0$ case, $1 - [dN/dz(z, \alpha)]/[dN/dz(z, \alpha = 0)]$, as a
function of $z$, for the values of $\Omega_0$ used in the upper panels. Here 
the lines correspond to $\alpha$ = 8, 4, and 2, in descending order at $z = 
1.5$.}
\end{figure}

\begin{figure}[p]
\psfig{file=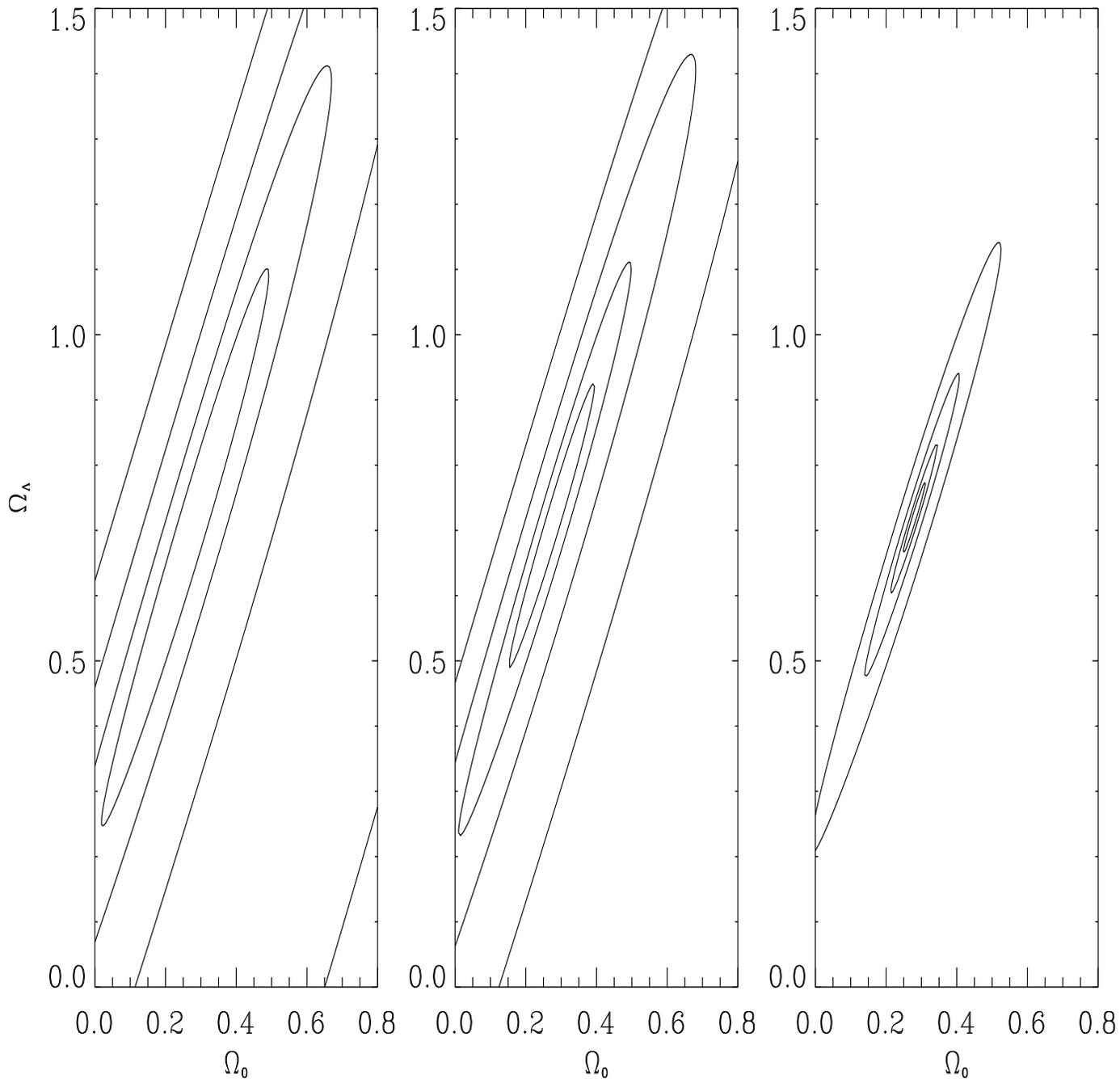,height=6.7in,width=7.0in,angle=90}
\vspace{0.5truein}
\caption{Contours of $N_\sigma$ = 1, 2, 4, and 8 for the constant $\Lambda$ 
model. Left panel is for anticipated DEEP data with worst case errors, center 
panel is for neutral case errors, and right panel is for best case errors. The 
fiducial model is spatially-flat with $\Omega_0 = 0.28$ and $\omegal = 0.72$.}
\end{figure}

\begin{figure}[p]
\psfig{file=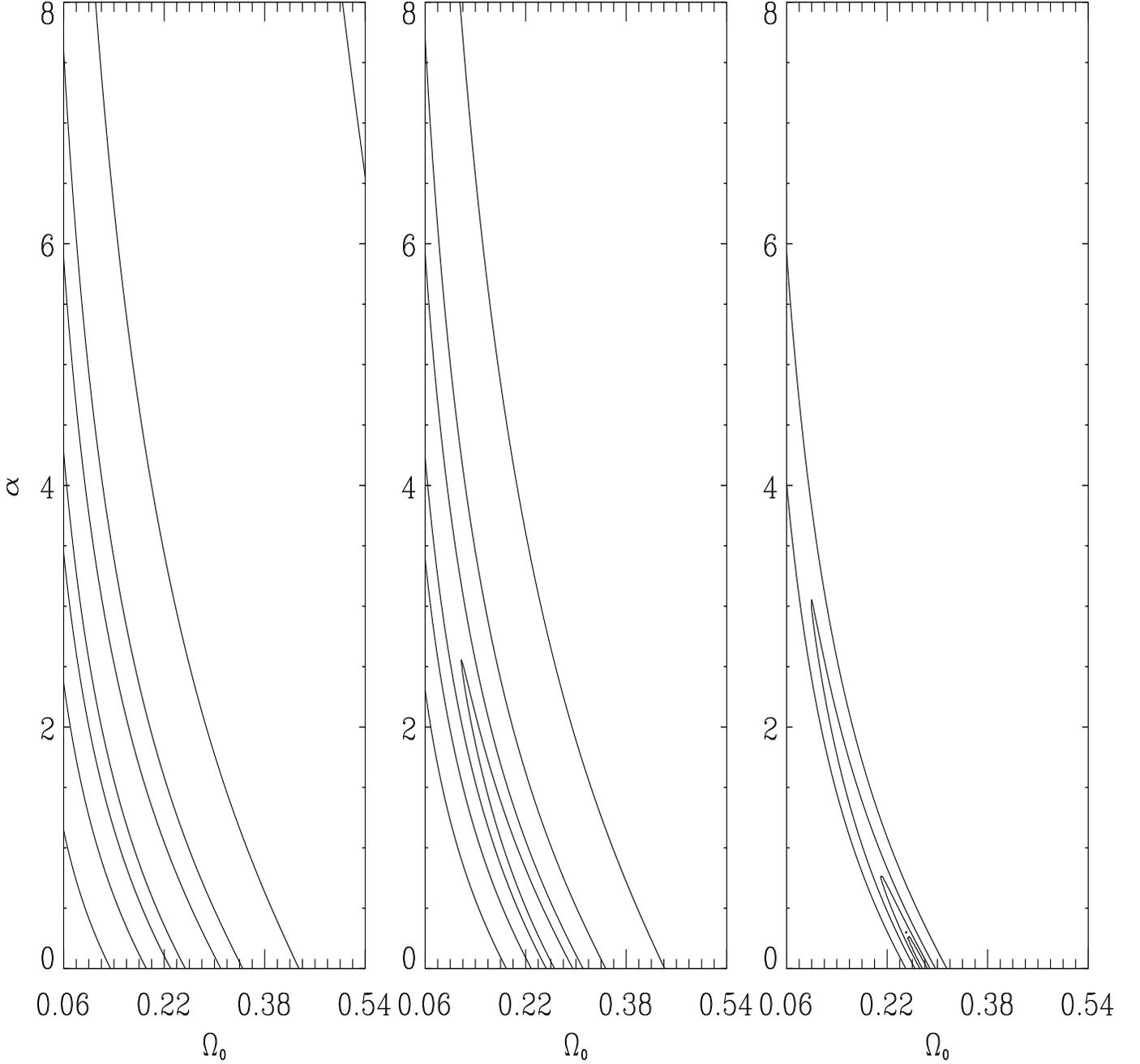,height=6.7in,width=7.0in,angle=90}
\vspace{0.75truein}
\caption{Contours of $N_\sigma$ = 1, 2, 4, and 8 for the spatially-flat 
time-variable $\Lambda$ model (PR). Left panel is for anticipated DEEP data 
with worst case errors, center panel is for neutral case errors, and right 
panel is for best case errors. The fiducial model has $\Omega_0 = 0.28$ and 
$\alpha = 0$ (and is thus a constant $\Lambda$ model with $\omegal = 0.72$; 
this is also the fiducial model used in Fig. 2).}
\end{figure}

\begin{figure}[p]
\psfig{file=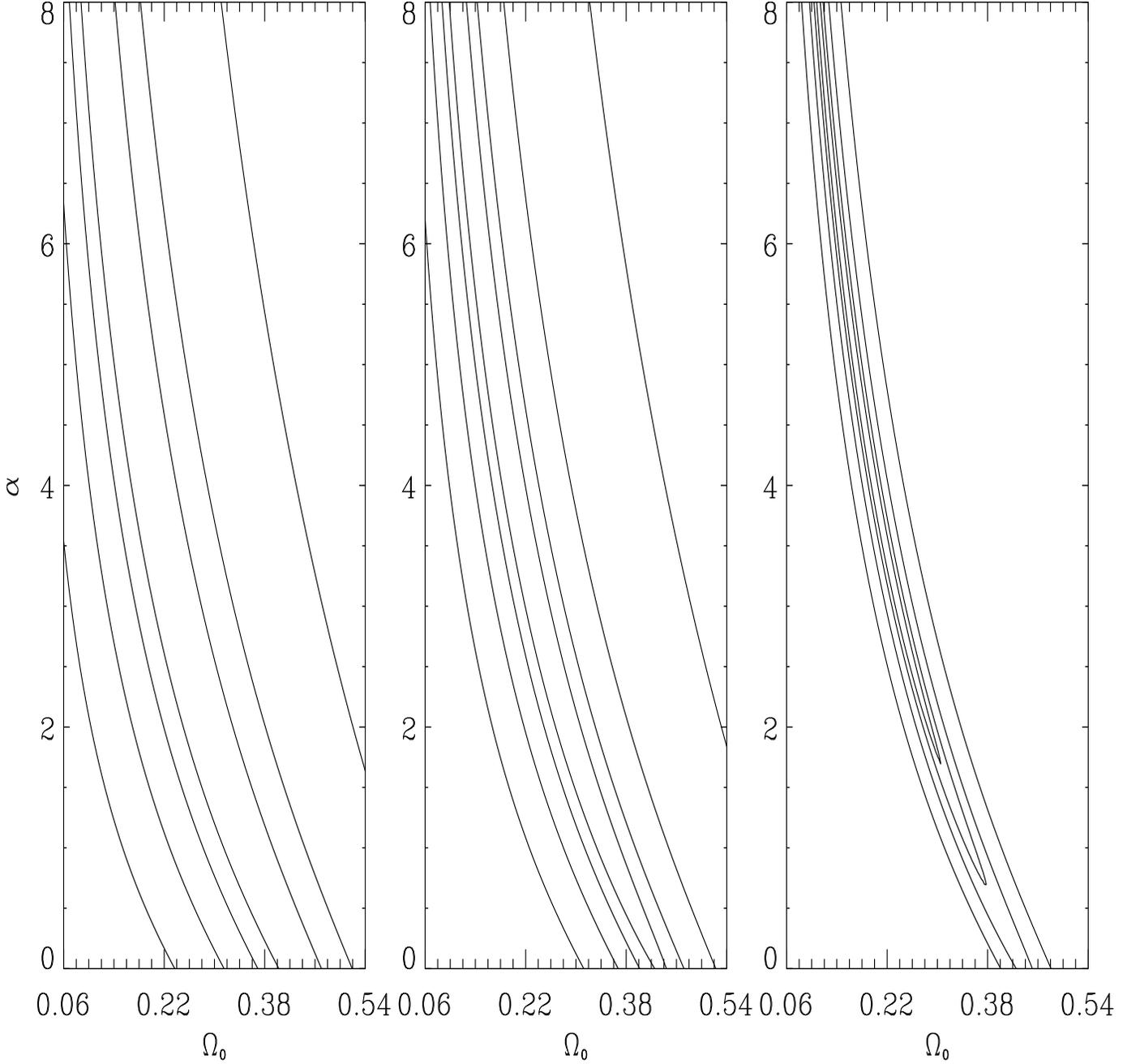,height=6.7in,width=7.0in,angle=90}
\vspace{0.75truein}
\caption{Contours of $N_\sigma$ = 1, 2, 4, and 8 for the spatially-flat 
time-variable $\Lambda$ model. Left panel is for anticipated DEEP data with 
worst case errors (part of the  $N_\sigma$ = 8 contour lies to the top and
right of the upper right hand corner of this panel), center panel is for 
neutral case errors, and right panel is for best case errors. The fiducial 
model has $\Omega_0 = 0.2$ and $\alpha = 4$.}
\end{figure}

\begin{figure}[p]
\psfig{file=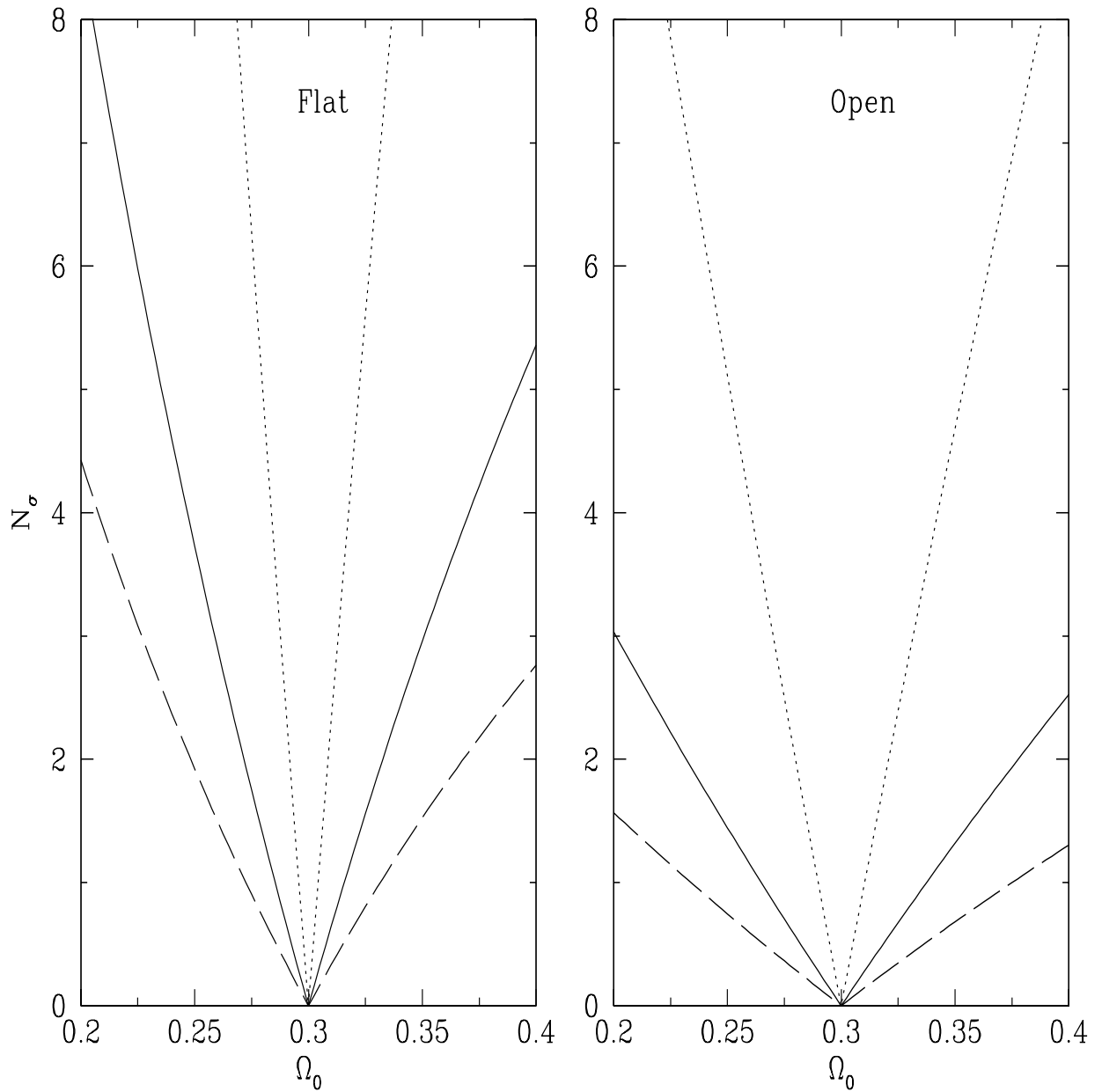,height=7.0in,width=7.0in,angle=270}
\caption{$N_\sigma (\Omega_0)$ for a flat model with a constant $\Lambda$ 
(left panel) and for an open model with no $\Lambda$ (right panel). In both 
cases the fiducial model has $\Omega_0 = 0.3$, with $\omegal = 0.7$ and 0, 
respectively. Solid lines are for neutral case DEEP errors while dotted 
(dashed) lines are for best (worst) case ones.}
\end{figure}

\begin{figure}[p]
\psfig{file=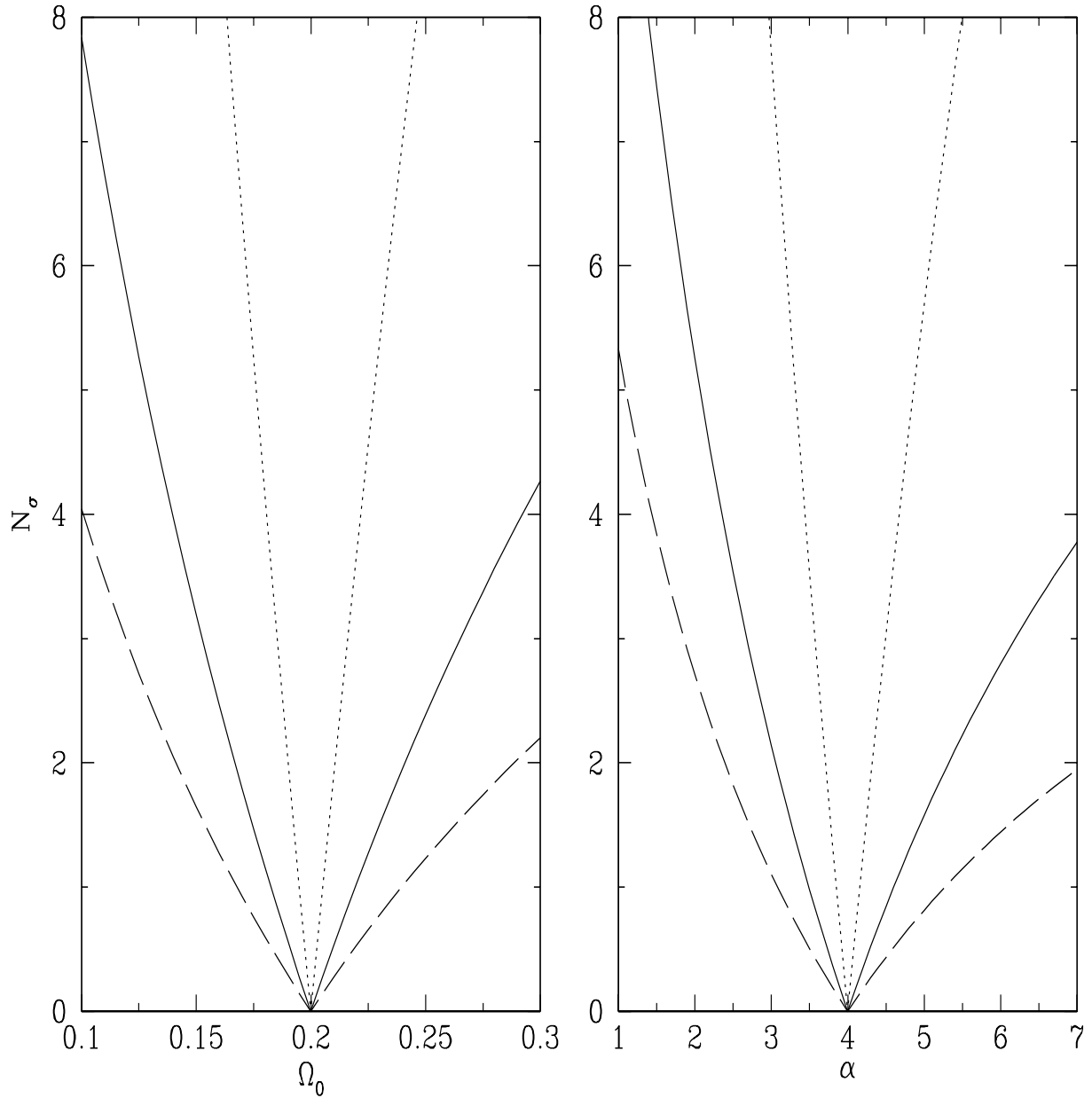,height=7.0in,width=7.0in,angle=270}
\caption{$N_\sigma (\Omega_0)$ (left panel) and
$N_\sigma(\alpha)$ (right panel) for the spatially-flat time-variable 
$\Lambda$ model (PR). In both cases the fiducial model has $\Omega_0 = 0.2$ 
and $\alpha = 4$.  Solid lines are for neutral case DEEP errors while dotted 
(dashed) lines are for best (worst) case ones.}
\end{figure}

\end{document}